\RequirePackage{fix-cm}
\documentclass[smallcondensed]{svjour3}     
\smartqed  
\usepackage{graphicx}
\usepackage{color}
\usepackage{enumitem}
\usepackage{amssymb}
\usepackage{multirow}
\begin{document}

\title{On Data Lake Architectures and Metadata Management}


\author{Pegdwend\'e Sawadogo \and
       J\'er{\^o}me Darmont 
}

\authorrunning{P. Sawadogo and J. Darmont} 

\institute{Pegdwend\'e Sawadogo \at
     Universit\'e de Lyon, Lyon 2, ERIC UR 3083\\
     \email{pegdwende.sawadogo@univ-lyon2.fr} 
\and 
J\'er\^ome Darmont \at
    Universit\'e de Lyon, Lyon 2, ERIC UR 3083\\
     \email{jerome.darmont@univ-lyon2.fr} 
}

\date{Received: date / Accepted: date}

\maketitle

\begin{abstract}
Over the past two decades, we have witnessed an exponential increase of data production in the world. So-called big data generally come from transactional systems, and even more so from the Internet of Things and social media. They are mainly characterized by volume, velocity, variety and veracity issues. 
Big data-related issues strongly challenge traditional data management and analysis systems. The concept of data lake was introduced to address them. A data lake is a large, raw data repository that stores and manages all company data 
bearing any format. 
However, the data lake concept remains ambiguous or fuzzy for many researchers and practitioners, who often confuse it with the Hadoop technology. Thus, we provide in this paper a comprehensive state of the art of the different approaches to data lake design. We particularly focus on data lake  architectures and metadata management, which are key issues in successful data lakes. We also discuss the pros and cons of data lakes and their design alternatives. 
\end{abstract}


\section*{Acknowledgement}
The research accounted for in this paper was funded by the Universit\'e Lumi\`ere Lyon 2 and the Auvergne-Rh\^one-Alpes Region through the COREL and AURA-PMI projects, respectively. The authors also sincerely thank the anonymous reviewers of this paper for their constructive comments and suggestions.

\section{Introduction}
\label{sec:intro}

The 21\textsuperscript{st} century is marked by an exponential growth of the amount of data produced in the world. This is notably induced by the fast development of the Internet of Things (IoT) and social media. Yet, while big data represent a tremendous opportunity for various 
organizations, 
they come in such volume, speed, heterogeneous sources and structures that they exceed the capabilities of traditional management systems for their collection, storage and processing in a reasonable time \cite{Miloslavskaya2016}.
A time-tested solution for big data management and processing is data warehousing. A data warehouse is indeed an integrated and historical storage system that is specifically designed to analyze data. However, while data warehouses are still relevant and very powerful for structured data, semi-structured and unstructured data induce great challenges for data warehouses. Yet, the majority of big data is unstructured \cite{Miloslavskaya2016}.
Thus, the concept of data lake was introduced to address big data issues, especially those induced by data variety.

A data lake is a very large data storage, management and analysis system that handles 
 any data format. It is currently quite popular and trendy both in the industry and academia.
Yet, the concept of data lake is not straightforward for everybody. 
A survey conducted in 2016 indeed revealed that 35\% of the respondents considered data lakes as a simple marketing label for a preexisting technology, i.e., Apache Hadoop \cite{Grosser2016}. 
Knowledge about the concept of the data lake has since evolved, but some misconceptions still exist,
presumably because most of data lakes design approaches are abstract sketches from the industry that provide few theoretical or implementation details \cite{Quix2018}. Therefore, a survey can be useful to give researchers and practitioners a better comprehension of the data lake concept and its design alternatives.  

\begin{figure}[hbt]
\centering
\includegraphics[width=12cm]{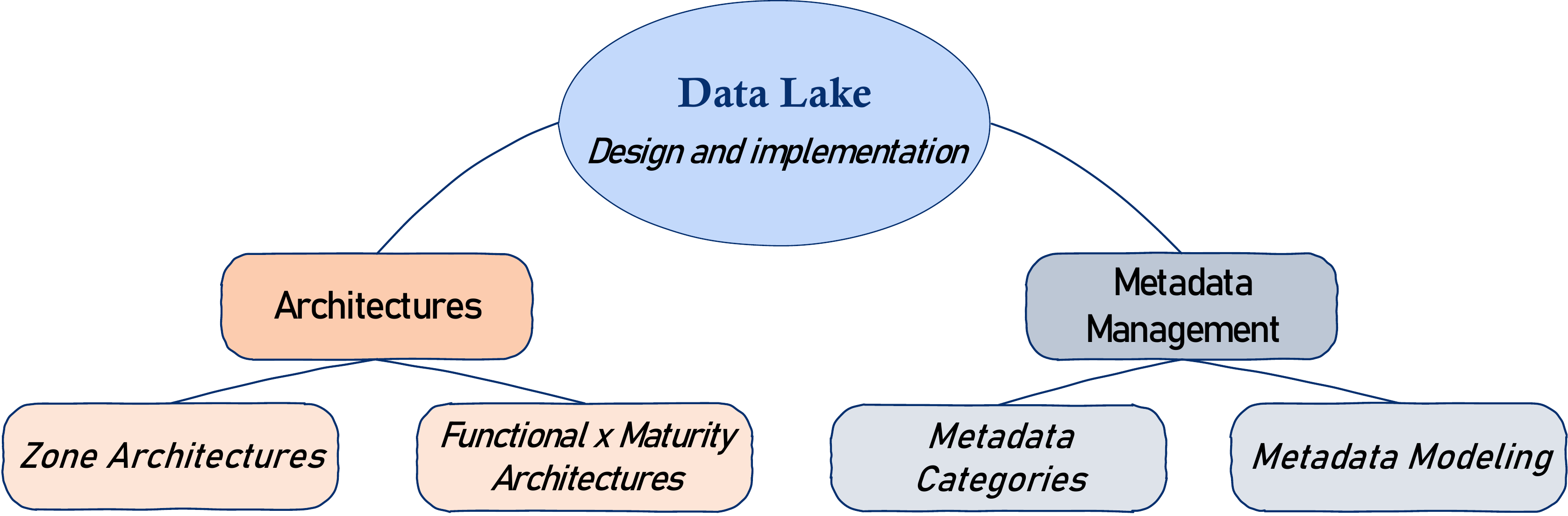} 
\caption{Issues addressed in this paper}
\label{fig:topics}
\end{figure}

To the best of our knowledge, the only literature reviews about data lakes are all quite brief 
and/or focused on a specific topic, e.g., data lake concepts and definitions \cite{Couto2019,Madera2016}, the technologies used for implementing data lakes \cite{Mathis2017} or data lakes inherent issues \cite{Giebler2019,Quix2018}. Admittedly, the report proposed in \cite{Russom2017} is quite extensive, but it adopts a purely industrial view.
Thus, we adopt in this paper a wider scope to propose a more comprehensive state of the art of the different approaches to design and exploit a data lake. We particularly focus on data lake architectures and metadata management, which lie at the base of any data lake project and are the most commonly cited issues in the literature (Figure~\ref{fig:topics}).

More precisely, we first review data lake definitions and complement the best existing one. Then, we investigate the architectures and technologies used for the implementation of data lakes, and propose a new typology of data lake architectures. Our second main focus is metadata management, which is a primordial issue to avoid turning a data lake into an inoperable, so-called data swamp. We notably classify data lake metadata and introduce the features that are necessary to achieve a full metadata system. We also discuss the pros and cons of data lakes. 

Eventually, note that we do not review other important topics, such as data ingestion, data governance and security in data lakes, because they are currently little addressed in the literature, but could still presumably be the subject of another full survey.

The remainder of this paper is organized as follows. In Section~\ref{sec:def}, we define the data lake concept. In Section~\ref{sec:archi}, we review data lake architectures and technologies to help users choose the right approach and tools.
In Section~\ref{sec:metadata}, we extensively review and discuss metadata management. 
Eventually, we recapitulate the pros and cons of data lakes in Section~\ref{sec:proscons} and conclude the paper in Section~\ref{sec:conclusion} with a mind map of the key concepts we introduce, as well as current open research issues.

\section{Data Lake Definitions}
\label{sec:def}

\subsection{Definitions from the Literature}
\label{sec:def.lit}

The concept of data lake was introduced by Dixon as a solution to perceived shortcomings of datamarts, which are business-specific subdivisions of data warehouses that allow only subsets of questions to be answered~\cite{Dixon2010}. 
In the literature, data lakes are also refered to as data reservoirs \cite{Chessell2014} and data hubs~\cite{Ganore2015,Laskowski2016}, although the terms data lake are the most frequent.
Dixon envisions a data lake as a large storage system for raw, heterogeneous data, fed by multiple data sources, and that allows users to explore, extract and analyze the data.

Subsequently, part of the literature considered data lakes as an equivalent to the Hadoop technology \cite{Fang2015,Ganore2015,OLeary2014}. According to this point of view, 
the concept of data lake refers to a methodology for using free or low-cost technologies, typically Hadoop, for storing, processing and exploring raw data within a company \cite{Fang2015}. 
The systematic association of data lakes to low cost technologies is becoming minority in the literature, as the data lake concept is now also associated with proprietary cloud solutions such as Azure or IBM~\cite{Madera2016,Sirosh2016} and various data management systems such as NoSQL solutions and multistores. However, it can still be viewed as a data-driven design pattern
for data management~\cite{Russom2017}.

More consensually, a data lake may be viewed as a central repository where data of all formats are stored without a strict schema, for future analyses~\cite{Couto2019,Khine2017,Mathis2017}. This definition is based on two key characteristics of data lakes: data variety and the schema-on-read approach, also known as late binding~\cite{Fang2015}, which implies that schema and data requirements are not fixed until data querying \cite{Khine2017,Maccioni2018,Stein2014}. This is the opposite to the schema-on-write approach used in data warehouses.

However, the variety/schema-on-read definition may be considered fuzzy because it gives little detail about the characteristics of a data lake. Thus, Madera and Laurent introduce a more complete definition where a data lake is a logical view of all data sources and datasets in their raw format, accessible by data scientists or statisticians for knowledge extraction~\cite{Madera2016}. 

More interestingly, this definition is complemented by a set of features that a data lake should include:
\begin{enumerate}
    \item data quality is provided by a set of metadata;
    \item the lake is controlled by data governance policy tools;
    \item usage of the lake is limited to statisticians and data scientists;
    \item the lake integrates data of all types and formats;
    \item the data lake has a logical and physical organization.
\end{enumerate}



\subsection{Discussion and New Definition}
\label{sec:def.disc}
Madera and Laurent's definition of data lakes is presumably the most precise, as it defines the requirements that a data lake must meet (Section~\ref{sec:def.lit}). 
However, some points in this definition are debatable. 
The authors indeed restrain the use of the lake to data specialists and, as a consequence, exclude business experts for security reasons. Yet, in our opinion, it is entirely possible to allow controlled access to this type of users through a navigation or analysis software layer.
        
Moreover, we do not share the vision of the data lake as a logical view over data sources, since some data sources may be external to an organization, and therefore to the data lake. Since Dixon explicitly states that lake data come from data sources~\cite{Dixon2010}, including data sources into the lake may therefore be considered contrary to the spirit of data lakes.

Finally, although quite complete, Madera and Laurent's definition omits an essential property of data lakes: 
scalability~\cite{Khine2017,Miloslavskaya2016}. Since a data lake is intended for big data storage and processing, it is indeed essential to address this issue. 
Thence, we amend Madera and Laurent's definition to bring it in line with our vision and introduce scalability~\cite{Sawadogo2019B}.

\begin{definition}
    A data lake is a scalable storage and analysis system for data of any type, retained in their native format and used \emph{mainly} by data specialists (statisticians, data scientists or analysts) for knowledge extraction. 
    Its characteristics include: 
    \begin{enumerate}
        \item a metadata catalog that enforces data quality;
        \item data governance policies and tools;
        \item accessibility to various kinds of users;
        \item integration of any type of data;
        \item a logical and physical organization;
        \item scalability in terms of storage and processing.
    \end{enumerate}
\end{definition}

\section{Data Lake Architectures and Technologies}
\label{sec:archi}
Existing reviews on data lake architectures commonly distinguish pond and zone architectures~\cite{Giebler2019,Ravat2019B}. However, this categorization may sometimes be fuzzy. Thus, we introduce in Section~\ref{sec:archi.internal} a new manner to classify data lakes architectures that we call Functional $\times$ Maturity. In addition, we present in Section~\ref{sec:archi.techno} a list of possible technologies to implement a data lake.
Eventually, we investigate in Section~\ref{sec:archi.global} how a data lake system can be associated with a data warehouse in an enterprise data architecture.

\subsection{Data Lake Architectures}
\label{sec:archi.internal}

\subsubsection{Zone Architectures}
\label{sec:archi.internal.literature}

\begin{figure}[hbt]
\centering
\includegraphics[width=10cm]{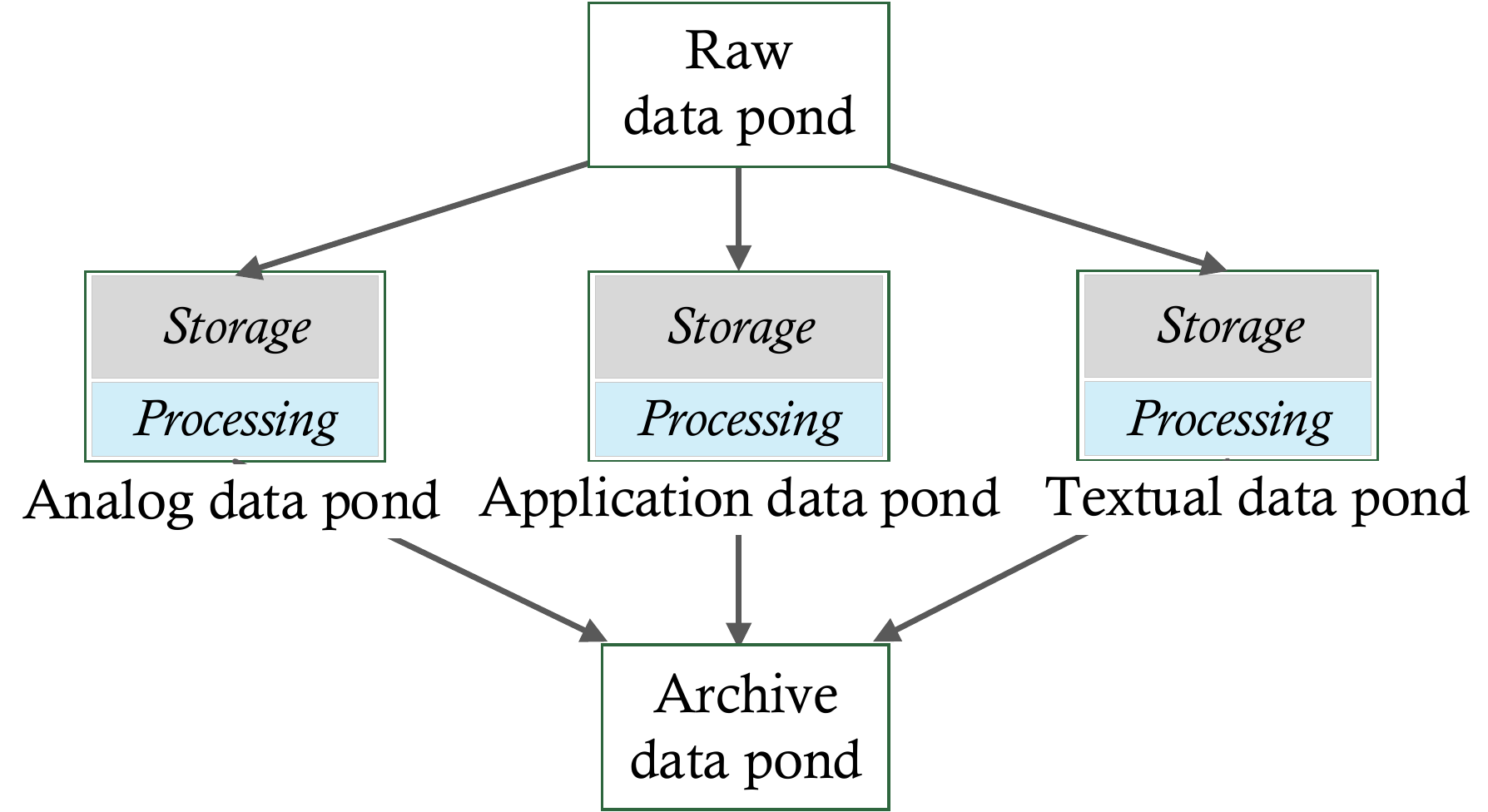} 
\caption{Data flow in a pond architecture}
\label{fig:data.ponds}
\end{figure}

\paragraph{Pond architecture}
Inmon designs a data lake as a set of data ponds~\cite{Inmon2016}. A data pond can be viewed as a subdivision of a data lake dealing with data of a specific type. According to Dixon's specifications, each data pond is associated with a specialized storage system, some specific data processing and conditioning (i.e., data transformation/preparation) and  a relevant analysis service. More precisely, Inmon identifies five data ponds (Figure~\ref{fig:data.ponds}). 

\begin{enumerate}
    \item The \textbf{raw data pond} deals with newly ingested, raw data. It is actually a transit zone, since data are then conditioned and transferred into another data pond, i.e., either the analog, application or textual data pond. The raw data pond, unlike the other ponds, is not associated with any metadata system. 
    
    \item Data stored in the \textbf{analog data pond} are characterized by a very high frequency of measurements, i.e., they come in with high velocity. Typically, semi-structured data from the IoT are processed in the analog data pond.
    
    \item Data ingested in the \textbf{application data pond} come from software applications, and are thus generally structured data from relational Database Management Systems (DBMSs). Such data are integrated, transformed and prepared for analysis; and Inmon actually considers that the application data pond is a data warehouse.
    
    \item The \textbf{textual data pond}  manages unstructured, textual data. It features a textual disambiguation process to ease textual data analysis.
    
    \item The purpose of the \textbf{archival data pond} is to save the data that are not actively used, but might still be needed in the future. Archived data may originate from the analog, application and textual data ponds. 
\end{enumerate}

\begin{figure}[hbt]
\centering
\includegraphics[width=12cm]{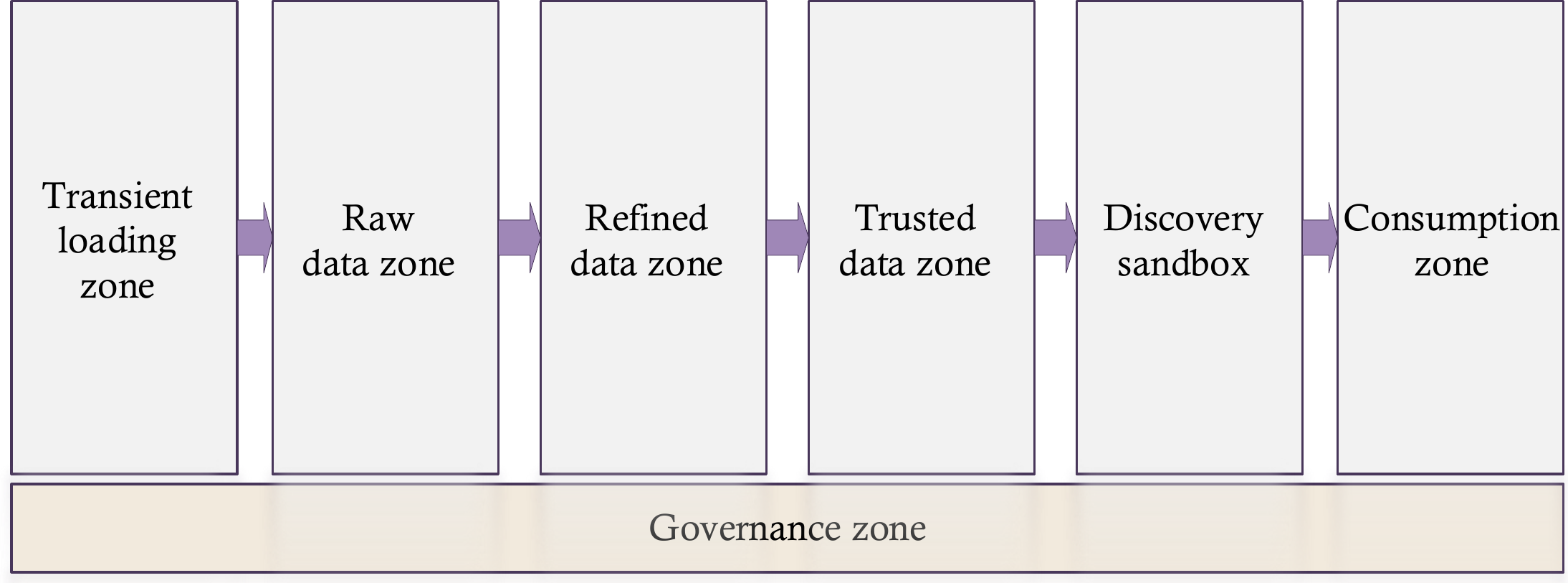} 
\caption{Zaloni's zone architecture~\cite{Laplante2016}}
\label{fig:zaloni}
\end{figure}

\paragraph{Zone architectures}
So-called zone architectures assign data to a zone according to their degree of refinement~\cite{Giebler2019}.  
For instance, Zaloni's data lake~\cite{Laplante2016} adopts a six-zone architecture (Figure~\ref{fig:zaloni}). 

\begin{enumerate}
    \item The \textbf{transient loading zone} deals with data under ingestion. Here, basic data quality checks are performed.
    
    \item The \textbf{raw data zone} handles data in near raw format coming from the transient zone. 
    
    \item The \textbf{trusted zone} is where data are transferred once standardized and cleansed. 
    
    \item From the trusted area, data move into the \textbf{discovery sandbox} where they can be accessed by data scientists through data wrangling or data discovery operations. 
    
    \item On top of the discovery sandbox, the \textbf{consumption zone}  allows business users to run ``what if'' scenarios through dashboard tools. 
    
    \item The \textbf{governance zone} finally allows to manage, monitor and govern metadata, data quality, a data catalog and security. 
\end{enumerate}

However, this is but one of several variants of zone architectures. Such architectures indeed generally differ in the number and characteristics of zones~\cite{Giebler2019}, e.g., some architectures include a transient zone~\cite{Laplante2016,Tharrington2017,Zikopoulos2015} while others do not~\cite{Hai2016,Ravat2019B}. 

A particular zone architecture often mentioned in the data lake literature is the lambda architecture~\cite{John2017,Mathis2017}. It indeed 
stands out 
since it includes two data processing zones: a batch processing zone for bulk data and a real-time processing zone for fast data from the IoT~\cite{John2017}. These two zones help
handling fast data as well as bulk data in an adapted and specialized way.

\paragraph{Discussion}
In both pond and zone architectures, data are pre-processed. Thus, analyses are quick and easy. However, this come at the cost of data loss in the pond architectures, since raw data are deleted when transferred to other ponds. The drawbacks of the many zone architectures depend on each variant. 
For example, in Zaloni's architecture~\cite{Laplante2016}, data flow across six areas, which may lead to multiple copies of the data and, therefore, difficulties in controlling data lineage.
In the Lamda architecture~\cite{John2017}, speed and batch processing components follow different paradigms. Thus, data scientists must handle two distinct logics for cross analyses~\cite{Mathis2017}, which makes 
data analysis harder, overall.

Moreover, the distinction of data lake architectures into pond and zone approaches is not so crisp in our opinion.  The pond architecture may indeed be considered as a variant of zone architecture, since data location depends on 
the refinement level of data, as in zone architectures.
In addition, some zone architectures 
include a global storage zone where raw and cleansed data are stored altogether~\cite{John2017,Quix2018}, which contradicts the definition of zone architectures, i.e., components depend on the degree of data refinement.

\subsubsection{Functional $\times$ Maturity Architectures}
\label{sec:archi.internal.new}

\begin{figure}[hbt]
\centering
\includegraphics[width=11cm]{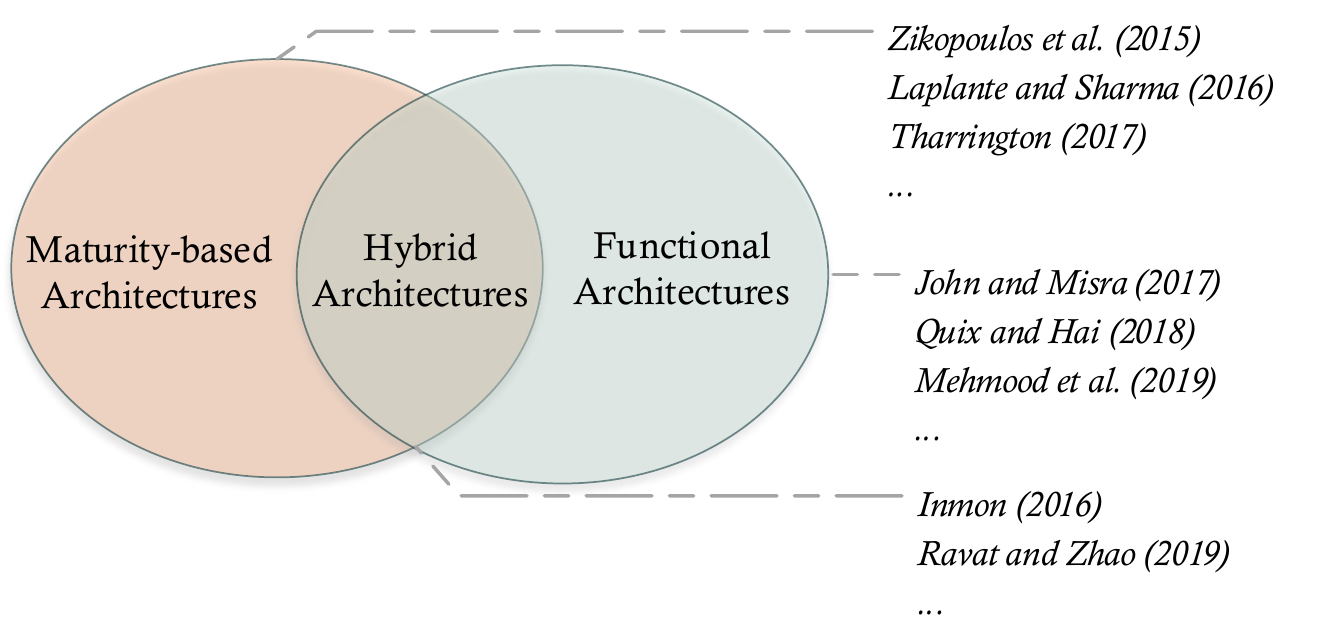} 
\caption{Architecture typology proposal}
\label{fig:new.archi.typology}
\end{figure}

To overcome the contradictions of the pond/zone categorization, we propose an alternative way to group data lake architectures 
regarding the type of criteria used to define components. As a result, we distinguish functional architectures, data maturity-based architectures and hybrid architectures (Figure~\ref{fig:new.archi.typology}).

\paragraph{Functional architectures}
follow some basic functions to define a lake's components. 
Data lake basic functions typically include~\cite{Laplante2016}: 
\begin{enumerate}
    \item a data ingestion function to connect with data sources;
    \item a data storage function to persist raw as well as refined data; 
    \item a data processing function; 
    \item a data access function to allow raw and refined data querying.
\end{enumerate}

Quix and Hai, as well as Mehmood et al., base their data lake architectures on these functions~\cite{Mehmood2019,Quix2018}. Similarly, John and Misra's lambda architecture \cite{John2017} may be considered as a functional architecture, since its components represent data lake functions such as storage, processing and serving. 

\paragraph{Data maturity-based architectures}
are data lake architectures where components are defined regarding data refinement level. In other words, it is constituted of most zone architectures. A good representative is Zaloni's data lake architecture~\cite{Laplante2016}, where common basic zones are a transient zone, a raw data zone, a trusted data zone and a refined data zone~\cite{Laplante2016,Tharrington2017,Zikopoulos2015}.

\begin{figure}[hbt]
\centering
\includegraphics[width=10cm]{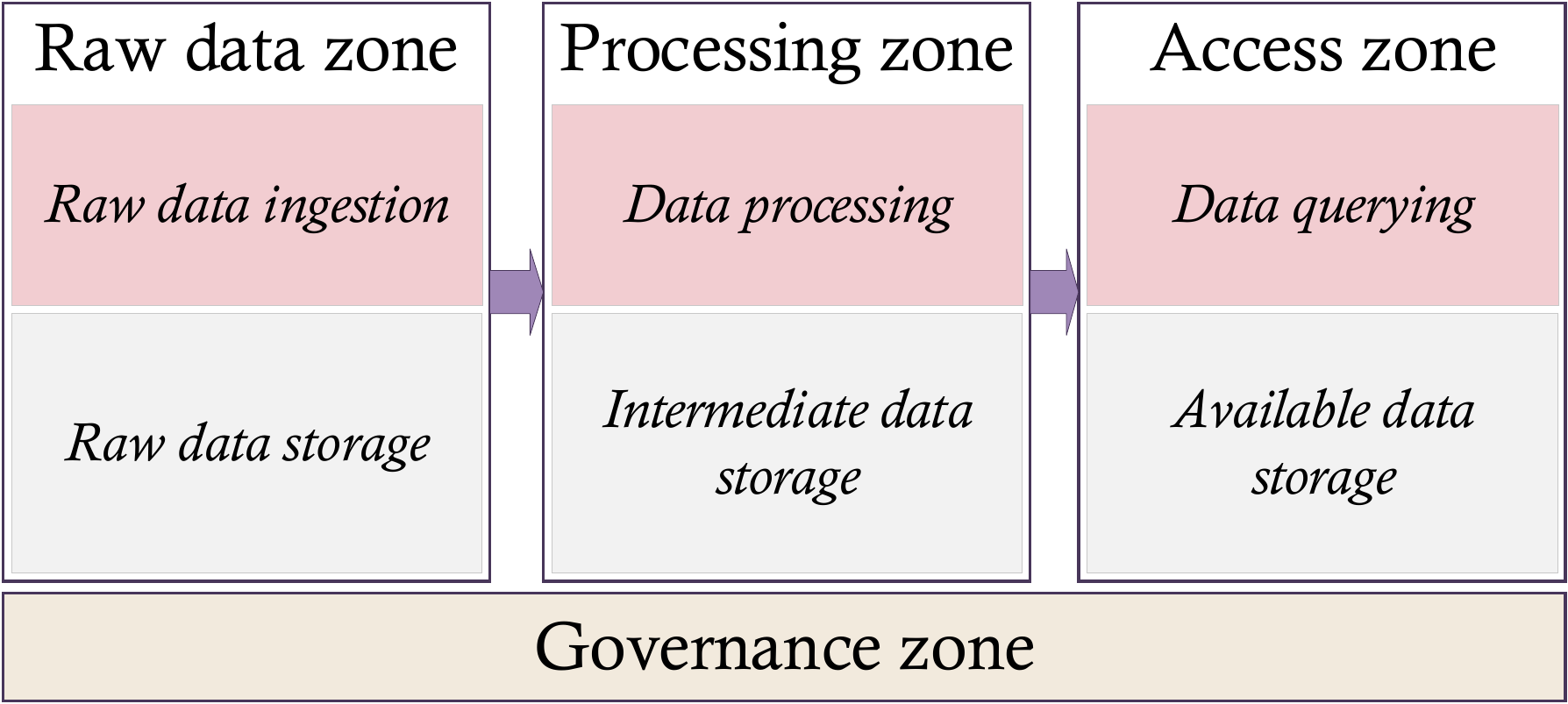} 
\caption{Ravat and Zhao's hybrid architecture~\cite{Ravat2019B}}
\label{fig:Ravat2019}
\end{figure}

\paragraph{Hybrid architectures}
are data lake architectures where the identified components depend on both data lake functions and data refinement. 
Inmon's pond architecture is actually a hybrid architecture~\cite{Inmon2016}. On one hand, it is a data maturity-based architecture, since raw data are managed in a special component, i.e., the raw data pond, while refined data are managed in other ponds, i.e., the textual, analog and application data ponds. 
But on the other hand, the pond architecture is also functional because Inmon's specifications consider some storage and process components distributed across data ponds (Figure~\ref{fig:data.ponds}). 
Ravat and Zhao also propose such an hybrid data lake architecture (Figure~\ref{fig:Ravat2019} \cite{Ravat2019B}).

\paragraph{Discussion}
Functional architectures have the advantage of clearly highlighting the functions to implement for a given data lake, which helps match easily with the required technologies. By contrast, data maturity-based architectures are useful to plan and organize the data lifecycle. Both approaches are thus limited, since they only focus on a unique point of view, while it is important in our opinion to take both functionality and data maturity into account when designing a data lake. 

In consequence, we advocate for hybrid approaches. 
However, existing hybrid architecture can still be improved. For instance, in Inmon's data pond approach, raw data are deleted once they are refined. This process may induce some data loss, which is contrary to the spirit of data lakes. 
In Ravat and Zhao's proposal, data access seems only possible for refined data. 
Such limitations hint that a more complete hybrid data lake architecture is still needed nowadays.

\subsection{Technologies for Data Lakes}
\label{sec:archi.techno}

Most data lake implementations are based on the Apache Hadoop ecosystem \cite{Couto2019,Khine2017}. Hadoop has indeed the advantage of providing both storage with the Hadoop Distributed File System (HDFS) and data processing tools via MapReduce or Spark. 
However, Hadoop is not the only suitable technology to implement a data lake. In this section, we go beyond Hadoop to review usable tools to implement data lake basic functions.

 \subsubsection{Data Ingestion}
\label{sec:archi.techno.ingest}
Ingestion technologies help physically transfer data from data sources into a data lake. A first category of tools includes software that iteratively collects data  through pre-designed and industrialized jobs. Most such tools are proposed by the Apache Foundation, and can also serve to aggregate, convert and clean data before ingestion. They include Flink and Samza (distributed stream processing frameworks), Flume (a Hadoop log transfer service),  Kafka (a framework providing real time data pipelines and stream processing applications) and Sqoop (a framework for data integration from SQL and NoSQL DBMSs into Hadoop)  \cite{John2017,Mathis2017,Suriarachchi2016}.%

A second category of data ingestion technologies is made of common data transfer tools and protocols (wget, rsync, FTP, HTTP, etc.), which are used by the data lake manager within data ingestion scripts. They have the key advantage to be readily available and widely understood  \cite{Terrizzano2015}. In a similar way, some Application  Programming  Interfaces (APIs) are available for data retrieval and transfer from the Web into a data lake. For instance,  CKAN  and Socrata provide APIs to access a catalogue of open data and associated metadata \cite{Terrizzano2015}.

 \subsubsection{Data Storage}
\label{sec:archi.techno.stor}
We distinguish two main approaches to store data in data lakes. The first way consists in using classic databases for storage. Some data lakes indeed use relational DBMSs such as MySQL, PostgreSQL or Oracle to store structured data \cite{Beheshti2017,Khine2017}. 
However, relational DBMSs are ill-adapted to semi-structured, and even more so to unstructured data. Thus, NoSQL (Not only SQL) DBMSs are usually used instead~\cite{Beheshti2017,Giebler2019,Khine2017}. 
Moreover, assuming that data variety is the norm in data lakes, a multi-paradigm storage system is particularly relevant~\cite{Nogueira2018}. Such so-called multistore systems manage  multiple DBMSs, each matching a specific storage need. 

The second main way to store data and the most used is HDFS storage (in about 75\% of data lakes~\cite{Russom2017}).  HDFS is a distributed storage system that offers a very high scalability and handles all types of data \cite{John2017}. Thus, it is well suited for schema-free and bulk storage that are needed for unstructured data. 
Another advantage of this technology is the distribution of data that allows high fault-tolerance. 
However, HDFS alone is not sufficient to handle all data formats, especially structured data. Thus, it should ideally be combined with relational and/or NoSQL DBMSs.

 \subsubsection{Data Processing}
\label{sec:archi.techno.proc}
In data lakes, data processing is very often performed with MapReduce \cite{Couto2019,John2017,Khine2017,Mathis2017,Stein2014,Suriarachchi2016}, a parallel data processing paradigm provided by Apache Hadoop. MapReduce is well-suited to very large data, but is less efficient for fast data because it works on disk~\cite{Tiao2018}. 
Thus, alternative processing frameworks are used, from which the most famous is Apache Spark. Spark works like MapReduce, but adopts a full in-memory approach instead of using the file system for storing intermediate results. Thence, Spark is  particularly suitable for real-time processing. Similarly, Apache Flink and Apache Storm are also suitable for real-time data processing~\cite{John2017,Khine2017,Mathis2017,Suriarachchi2016,Tiao2018}.
Nevertheless, these two approaches can be simultaneously implemented in a data lake, with MapReduce being dedicated to voluminous data and stream-processing engines to velocious data \cite{John2017,Suriarachchi2016}.

 \subsubsection{Data Access}
\label{sec:archi.techno.access}
  In data lakes, data may be accessed through classical query languages such as SQL for relational DBMSs, JSONiq for MongoDB, XQuery for XML DBMSs or SPARQL for RDF resources  \cite{Farid2016,Fauduet2010,Hai2016,Laskowski2016,Pathirana2015}. 
However, this does not allow simultaneously querying across heterogeneous databases, while data lakes do store heterogeneous data, and thus typically require heterogeneous storage systems. 

One solution to this issue is to adopt the query techniques from multistores (Section~\ref{sec:archi.techno.stor}) \cite{Nogueira2018}. For example, Spark SQL and SQL++ may be used to query both relational DBMSs and semi-structured data in JSON format. In addition, the Scalable Query Rewriting Engine (SQRE) handles graph databases~\cite{Hai2018}. Finally, CloudMdsQL also helps simultaneously query multiple relational and NoSQL DBMSs \cite{Leclercq2018}. 
Quite similarly, Apache Phoenix can be used to automatically convert SQL queries into a NoSQL  query language, for example. Apache Drill allows joining data from multiple storage systems \cite{Beheshti2017}. 
Data stored in HDFS can also be accessed using Apache Pig \cite{John2017}. 

Eventually, business users, who require interactive and user-friendly tools for data reporting and visualization tasks, widely use dashboard services such as Microsoft Power BI and Tableau over data lakes~\cite{Couto2019,Russom2017}.
 
\subsection{Combining Data Lakes and Data Warehouses}
\label{sec:archi.global}

There are in the literature two main approaches to combine a data lake and a data warehouse in a global data management system. The first approach pertains to using a data lake as the data source of a data warehouse (Section~\ref{subsec:dltodw}). The second considers data warehouses as components of data lakes (Section~\ref{subsec:dwindl}).

\subsubsection{Data Lake Sourcing a Data Warehouse}
\label{subsec:dltodw}

This approach aims to take advantage of the specific characteristics of both data lakes and data warehouses. Since data lakes allow an easier and cheaper storage of large amount of raw data, 
they can be considered as staging areas or Operational Data Stores (ODSs) \cite{Fang2015,Russom2017}, i.e., intermediary data stores ahead of data warehouses that gather operational data from several sources before the ETL process takes place.

\begin{figure}[hbt]
\centering
\includegraphics[width=8.5cm]{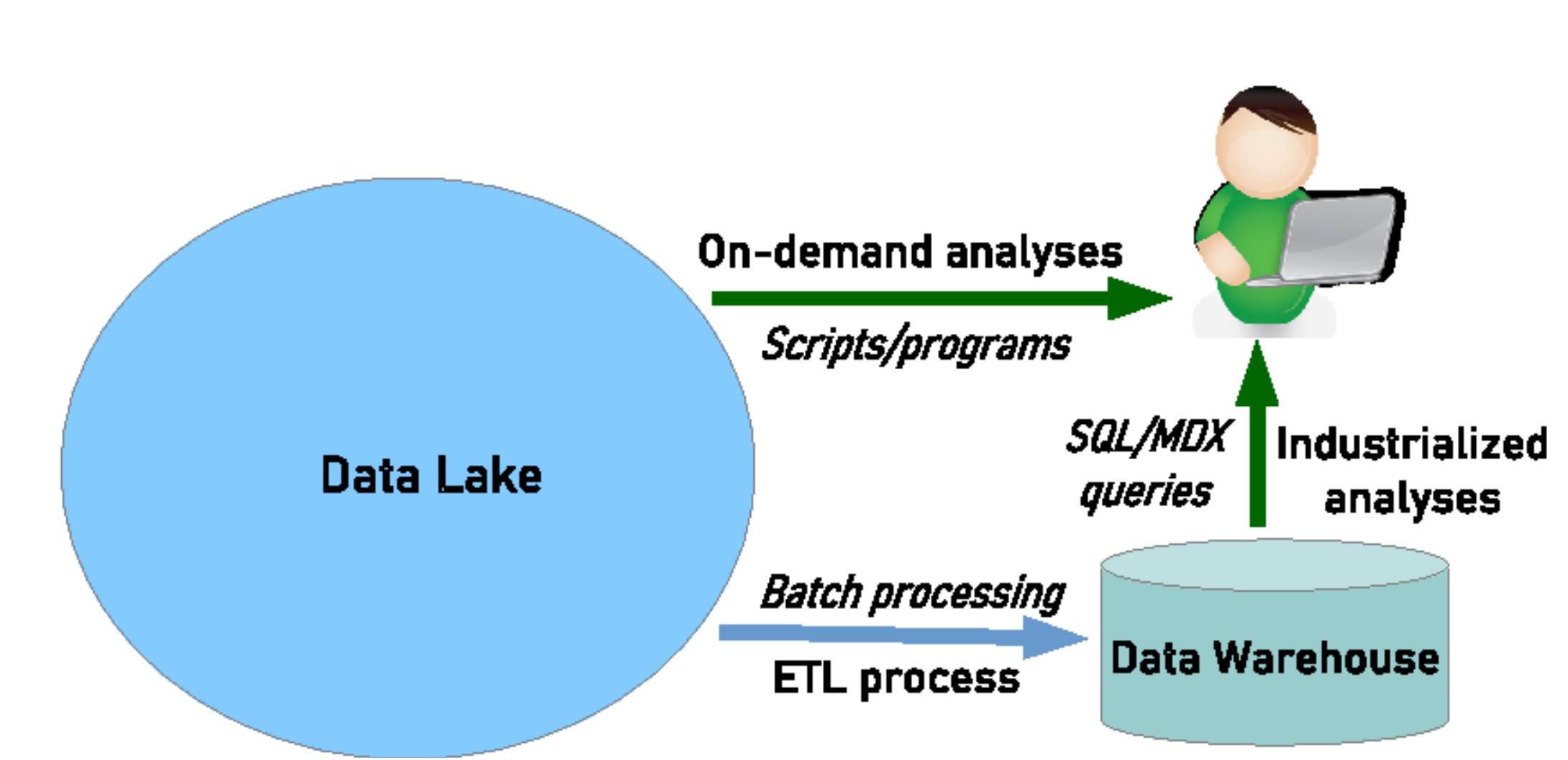} 
\caption{Data lake and data warehouse architecture}
\label{fig:dltodw}
\end{figure}

With a data lake sourcing a data warehouse, possibly with semi-structured data, industrialized OLAP analyses are possible over the lake's data, while on-demand, ad-hoc analyses are still possible 
directly from the data lake (Figure~\ref{fig:dltodw}). 

\subsubsection{Data Warehouse within a Data Lake}
\label{subsec:dwindl}
As detailed in Section~\ref{sec:archi.internal.literature}, Inmon proposes an architecture based on a subdivision of data lakes into so-called data ponds~\cite{Inmon2016}. 
For Inmon, structured data ponds sourced from operational applications are, plain and simple, data warehouses. 
Thus, this approach acts on a conception of data lakes as extensions of data warehouses.


\subsubsection{Discussion}
\label{subsec:discdwindl}

When a data lake sources a data warehouse (Section~\ref{subsec:dltodw}), there is a clear functional separation, as data warehouses and data lakes are specialized in industrialized and on-demand analyses, respectively. However, this comes with a data siloing issue.

By contrast, the data siloing syndrome can be reduced in Inmon's approach (Section~\ref{subsec:dwindl}), as all data are managed and processed in a unique global platform. Hence, diverse data can easily be combined through cross-reference analyses, which would be impossible if data were managed separately. 
In addition, building a data warehouse inside a global data lake may improve data lifecycle control. That is, it should be easier to track, and thus to reproduce processes applied to the data that are ingested in the data warehouse, via the data lake's tracking system.

\section{Metadata Management in Data Lakes}
\label{sec:metadata}

Data ingested in data lakes bear no explicit schema~\cite{Miloslavskaya2016}, which can easily turn a data lake into a data swamp in the absence of an efficient metadata system~\cite{Suriarachchi2016}. Thence, metadata management plays an essential role in data lakes~\cite{Laskowski2016,Khine2017}. 
In this section, we detail the metadata management techniques used in data lakes. First, we identify the metadata that are relevant to data lakes. 
Then, we review how metadata can be organized.  We also investigate metadata extraction tools and techniques.
Finally, we provide an inventory of desirable features in metadata systems. 

\subsection{Metadata Categories}
\label{sec:metadata.categ}

We identify in the literature two main typologies of metadata dedicated to data lakes. The first one distinguishes functional metadata, while the
second classifies metadata with respect to structural metadata types. 

\subsubsection{Functional Metadata}
\label{sec:metadata.categ.bot}

Oram introduces a metadata classification in three categories, with respect to the way they are gathered~\cite{Oram2015}. 
\begin{enumerate}
\item \textbf{Business metadata} are defined as the set of descriptions that make the data more understandable and define business rules. More concretely, these are typically data field names and integrity constraints. Such metadata are usually defined by business users at the data ingestion stage.

\item \textbf{Operational metadata} are information automatically generated during data processing. They include descriptions of the source and target data, e.g., data location, file size, number of records, etc., as well as process information. 

\item \textbf{Technical metadata} express how data are represented, including data format (e.g., raw text, JPEG image, JSON document, etc.), structure or schema. The data structure consists in characteristics such as names, types, lengths, etc. They are commonly obtained from a DBMS for structured data, or via custom techniques during the data maturation stage. 
\end{enumerate}

\begin{figure}[hbt]
\centering
\includegraphics[width=8cm]{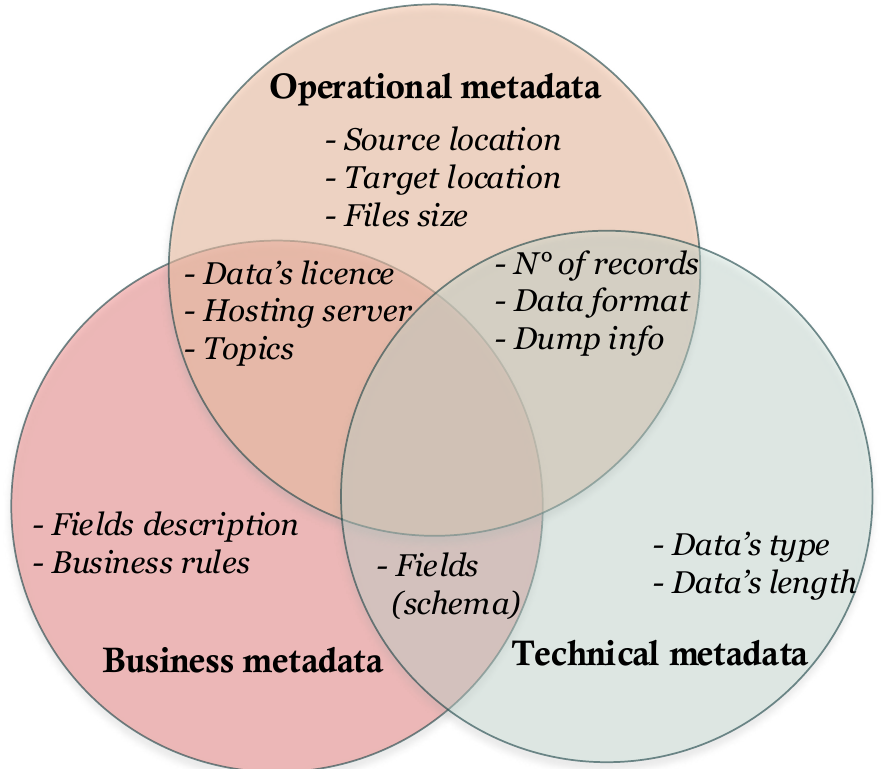}
\caption{Functional metadata model~\cite{Diamantini2018}}
\label{fig:BOTMetadata}
\end{figure}

Diamantini et al. enhance this typology with a generic metadata model~\cite{Diamantini2018} and show that business, operational and technical metadata sometimes intersect. For instance, data fields relate both to business and technical metadata, since they are defined in data schemas by business users. Similarly, data formats may be considered as both technical and operational metadata, and so on (Figure~\ref{fig:BOTMetadata}).

\subsubsection{Structural Metadata}
\label{sec:metadata.categ.int}

In this classification, Sawadogo et al. categorize metadata with respect to the ``objects'' they relate to~\cite{Sawadogo2019B}. The notion of object may be viewed as a generalization of the dataset concept \cite{Maccioni2018}, i.e., an object may be a relational or spreadsheet table in a structured data context, or a simple document (e.g., XML document, image file, video file, textual document, etc.) in a   semi-structured or unstructured data context. Thence, we use the term ``object'' in the remainder of this paper.

\paragraph{Intra-object metadata} belong to a set of characteristics associated with single objects in the lake.  They are subdivided into four main subcategories.

\begin{enumerate}
\item \textbf{Properties} provide an object's general description. They are generally retrieved from the filesystem as key-value pairs, e.g., file name and size, location, date of last modification, etc.

\item \textbf{Previsualization and summary metadata} aim to provide an overview of the content or structure of an object. For instance, metadata can be extracted data schemas 
for structured and semi-structured data, or wordclouds for textual data. 

\item \textbf{Version and representation metadata} are made of altered data. 
When a new data object $o'$ is generated from existing object $o$ in the data lake, $o'$ may be considered as  metadata for $o$. Version metadata are obtained through data updates, while representation metadata come from data refining operations. For instance, a refining operation may consist of vectorizing a textual document into a bag-of-words for further automatic processing. 

\item \textbf{Semantic metadata} involve annotations that describe the meaning of data in an object. They include such information as title, description, categorization, descriptive tags, etc. They often allow data linking.
Semantic metadata can be either generated using semantic resources such as ontologies, or manually added by business users \cite{Hai2016,Quix2016}. 
\end{enumerate}

\paragraph{Inter-object metadata} represent links between two or more objects. They are subdivided into three categories. 
\begin{enumerate}
\item \textbf{Object groupings} organize objects into collections. Any object may be associated with several collections. Such links can be automatically deduced from some intra-object metadata such as tags, data format, language, owner, etc.

\item \textbf{Similarity links} express the strength of likeness between  objects. They are obtained via common or custom similarity measures. For instance, \cite{Maccioni2018} define the affinity and joinability measures to express the similarity between semi-structured objects. 

\item \textbf{Parenthood links} aim to save data lineage, i.e., when a new object is created from the combination of several others, these metadata 
record the process. Parenthood links are thus automatically generated during data joins. 
\end{enumerate}

\paragraph{Global metadata} are data structures that provide a context layer to make data processing and analysis  easier. Global metadata are not directly associated with any specific object, but potentially concern the entire lake. There are three subcategories of global metadata.

\begin{enumerate}
\item \textbf{Semantic resources} are knowledge bases such as ontologies, taxonomies, thesauri, etc., which notably help enhance analyses. For instance, an ontology can be used to automatically extend a term-based query with equivalent terms. Semantic resources are generally obtained from the Internet or manually built. 

\item \textbf{Indexes} (including inverted indexes) enhance term-based or pattern-based data retrieval. They are automatically built and enriched by an indexing system.

\item \textbf{Logs} track user interactions with the data lake, which can be simple, e.g., user connection or disconnection, or more complex, e.g., a job running. 
\end{enumerate}

\subsubsection{Discussion}
\label{sec:metadata.categ.disc}

Oram's metadata classification is 
the most cited, especially in the industrial literature~\cite{Diamantini2018,Laplante2016,Ravat2019A,Russom2017}, 
presumably because it is inspired from metadata categories from data warehouses~\cite{Ravat2019B}. Thus, its adoption seems easier and more natural for practitioners who are already working with it.

Yet, we favor the second metadata classification, because it includes most of the features defined by Oram's. Business metadata are indeed comparable to semantic metadata. Operational metadata may be considered as logs and technical metadata are equivalent to previsualization metadata. 
Hence, the structural metadata categorization can be considered as an extension, as well as a generalization, of the functional metadata classification.

Moreover, Oram's classification is quite fuzzy when applied in the context of data lakes. Diamantini et al. indeed show that functional metadata intersect  (Section~\ref{sec:metadata.categ.bot})~\cite{Diamantini2018}. Therefore, practitioners who do not know this typology may be confused when using it to identify and organize metadata in a data lake. 

Table~\ref{tab:metadata.categ} summarizes commonalities and differences between the two metadata categorizations presented above. The comparison addresses the type of information both inventories provide.

\begin{table*}[hbt]
\centering
\caption{Comparison of Oram's and Sawadogo et al's metadata categories}
\begin{tabular} {l c c }
\hline
 \textbf{Type of information} &  \textbf{Functional metadata} &  \textbf{Structural metadata}  \\
  \hline  
   \hline
   Basic characteristics of data & \multirow{2}{*}{\checkmark} & \multirow{2}{*}{\checkmark} \\
    (size, format, etc.) & & \\
     
    \hline
    Data semantics   &  \multirow{2}{*}{\checkmark} & \multirow{2}{*}{\checkmark} \\
    (tags, descriptions, etc.) & & \\
     \hline
   Data history  &   \checkmark  &  \checkmark  \\
     \hline
     Data linkage &  & \checkmark  \\
     \hline
    User interactions &   & \checkmark  \\
    \hline
\end{tabular}
\label{tab:metadata.categ}
\end{table*}

\subsection{Metadata Modeling}
\label{sec:metadata.model}

There are in the literature two main approaches to represent a data lake's metadata system. The first, most common approach, adopts a graph view, while the second exploits data vault modeling.

\subsubsection{Graph Models}
\label{sec:metadata.model.graph}
Most models that manage data lake metadata systems are based on a graph approach. We identify three main subcategories of graph-based metadata models with respect to the main features they target. 

\begin{figure*}[hbt]
\centering
\includegraphics[width=0.50\textwidth]{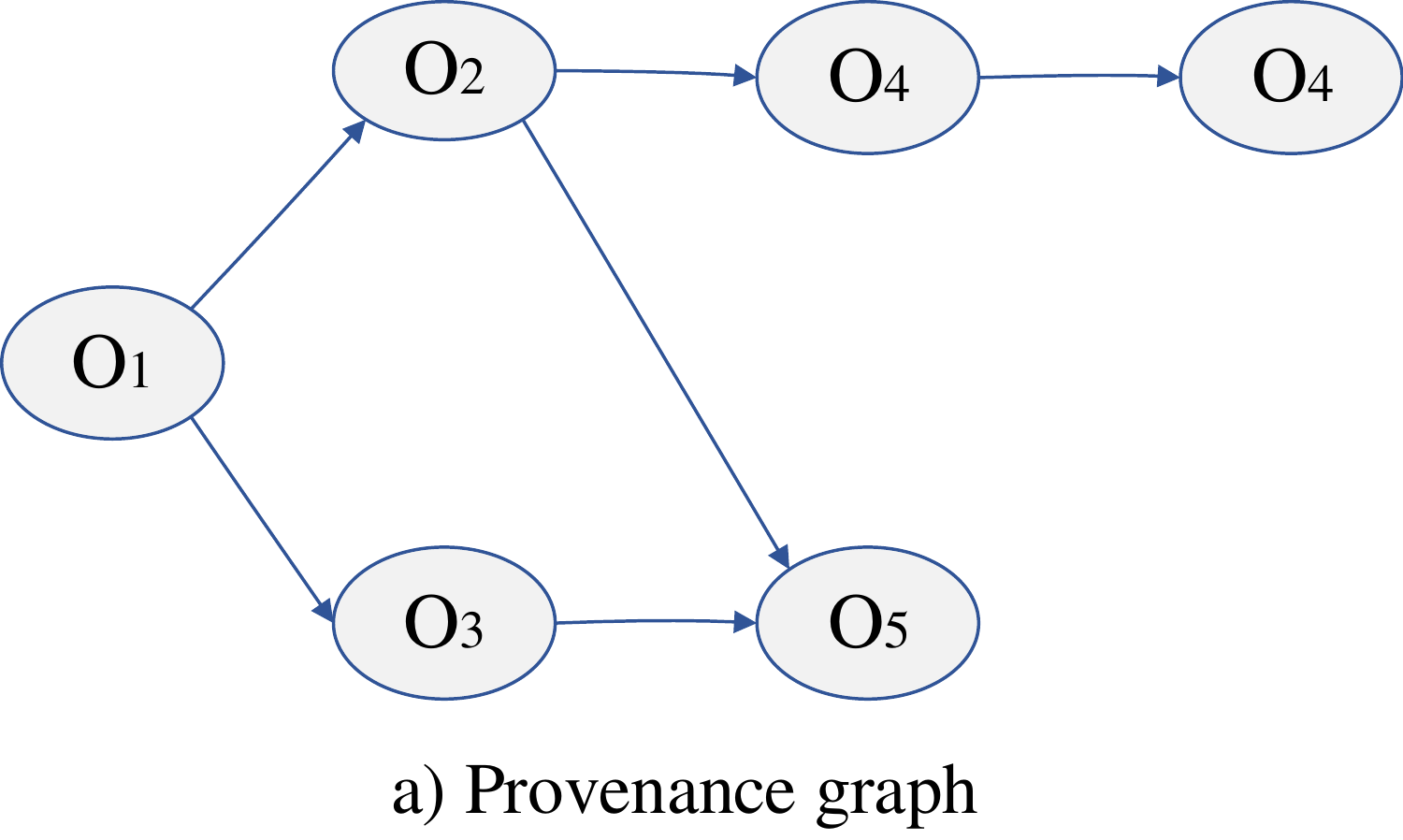}
\qquad \qquad 
\includegraphics[width=0.33\textwidth]{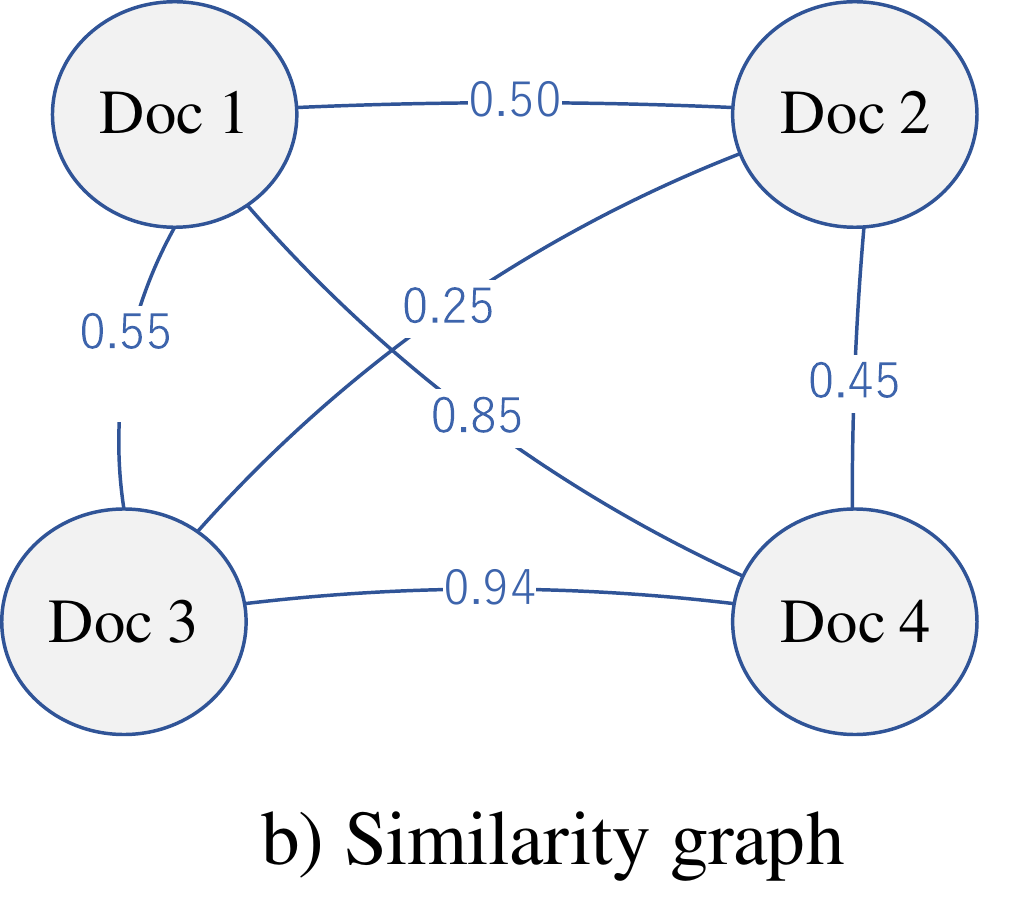}
\caption{Sample graph-based metadata models}
\label{fig:graph.models}
\end{figure*}

\paragraph{Data provenance-centered graph models} mostly manage metadata tracing, i.e., the information about activities, data objects and users who interact with a specific object \cite{Suriarachchi2016}. In other words, they track the pedigree of data objects \cite{Halevy2016B}.
Provenance representations are usually built using a directed acyclic graph (DAG) where nodes represent entities such as users, roles or objects \cite{Beheshti2017,Hellerstein2017}. Edges are used to express and describe interactions between entities, e.g., through a simple timestamp, activity type (read, create, modify) \cite{Beheshti2017}, system status (CPU, RAM, bandwith) \cite{Suriarachchi2016} or even the script used~\cite{Hellerstein2017}. For instance Figure~\ref{fig:graph.models}-a shows a basic provenance model with nodes representing data objects and edges symbolizing operations.
Data provenance tracking helps ensure the traceability and repeatability of  processes
in data lakes. Thus, provenance metadata can be used to understand, explain and repair inconsistencies in the data~\cite{Beheshti2017}. They may also serve to protect sensitive data, by detecting intrusions~\cite{Suriarachchi2016}.

\paragraph{Similarity-centered graph models} describe the metadata system as an undirected graph where nodes are data objects and edges express a similarity between objects. Such a similarity can be specified either through a weighted or unweighted edge. Weighted edges show the similarity strength, when a formal similarity measure is used, e.g., affinity and joinability~\cite{Maccioni2018} (Figure~\ref{fig:graph.models}-b). In contrast, unweighted edges serve to simply detect whether two objects are connected~\cite{Farrugia2016}.
Such a graph design allows network analyses over a data lake~\cite{Farrugia2016}, e.g., discovering communities or calculating the centrality of nodes, and thus their importance in the lake.
Another use of data similarity may be to automatically recommend to lake users some data related to the data they currently observe~\cite{Maccioni2018}.

\paragraph{Composition-centered graph models} help decompose each data object into several inherent elements. The lake is viewed as a DAG where nodes represent objects or attributes, e.g., columns, tags, etc., and edges from any node $A$ to any node $B$ express the constraint $B \subseteq A$ \cite{Diamantini2018,Halevy2016B,Nargesian2018}. 
This organization helps users navigate through the data~\cite{Nargesian2018}. It can also be used as a basis to detect connections between objects. For instance, \cite{Diamantini2018} used a simple string measure to detect links between heterogeneous objects by comparing their respective tags.


\subsubsection{Data Vault}
\label{sec:metadata.model.vault}

\begin{figure}[hbt]
\centering
\includegraphics[width=12cm]{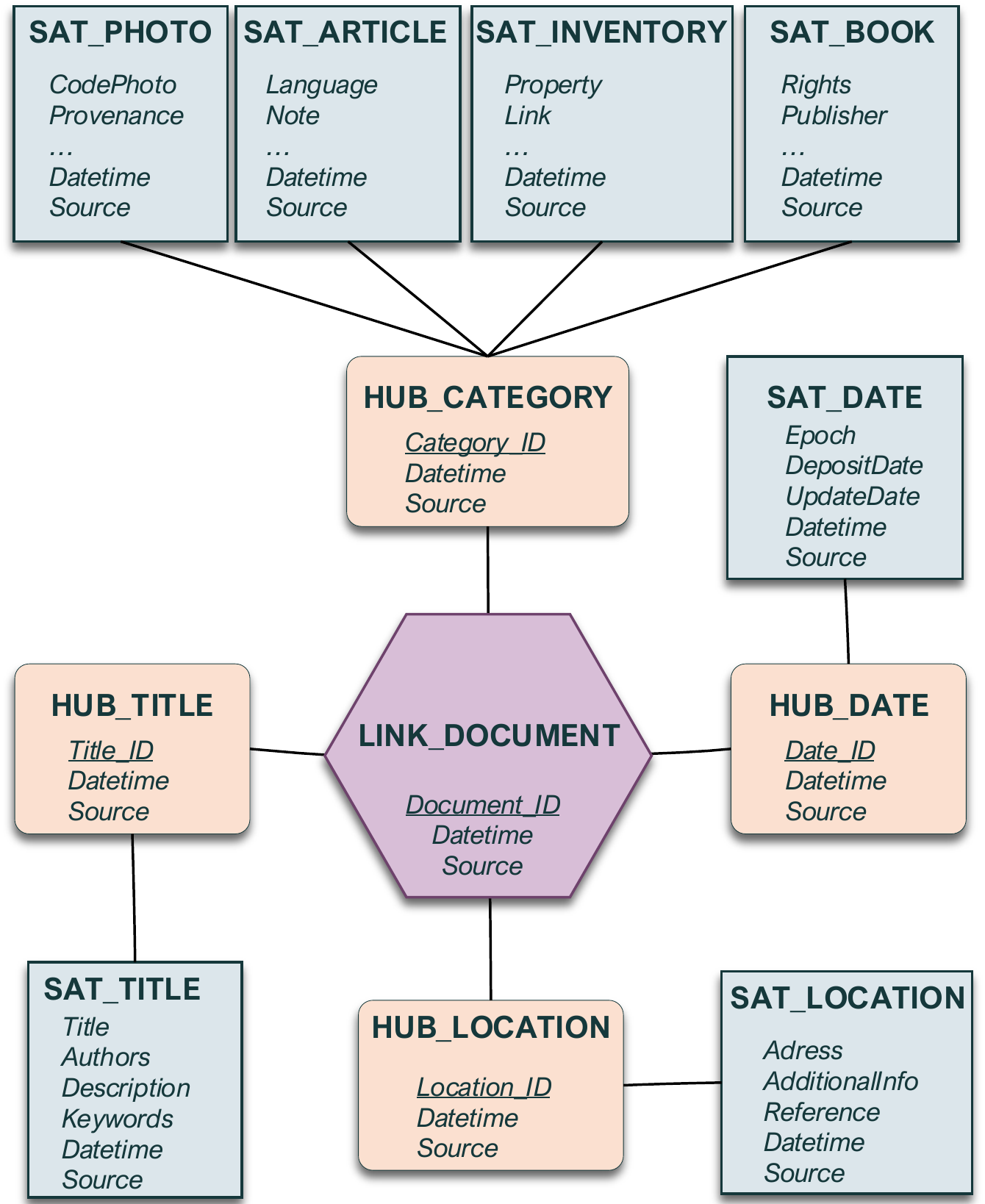} 
\caption{Sample metadata vault model \cite{Nogueira2018}}
\label{fig:data.vault}
\end{figure}

A data lake aims at ingesting new data possibly bearing various structures. Thus, its metadata system needs to be flexible to easily tackle new data schemas. Nogueira et al. propose the use of a data vault to address this issue~\cite{Nogueira2018}. Data vaults are indeed alternative logical models to data warehouse star schemas that, unlike star schemas, 
allow easy schema evolution \cite{Linstedt2011}. 
Data vault modeling involves three types of entities~\cite{Hultgren2016}.
\begin{enumerate}
    \item A \textbf{hub} represents a business concept, e.g., customer, vendor, sale or product in a business decision system.
    \item A \textbf{link} represents a relationship between two or more hubs.
    \item \textbf{Satellites} contain descriptive information associated with a hub or a link. Each satellite is attached to a unique hub or link.  In contrast, links or hubs may be associated with any number of satellites.
\end{enumerate}
In Nogueira et al.'s proposal, metadata common to all objects, e.g., title, category, date and location, are stored in hubs; while metadata specific to some objects only, e.g., language for textual documents or publisher for books, are stored in satellites (Figure~\ref{fig:data.vault}). Moreover, any new type of object would have its specific metadata stored in a new satellite.
\label{subsec:DVMod}

\subsubsection{Discussion}
\label{sec:metadata.model.disc}  
Data vault modeling is seldom associated with data lakes in the literature, presumably because it is primarily associated with data warehouses. Yet, this approach ensures metadata schema evolutivity, which is required to build an efficient data lake. Another advantage of data vault modeling is that, unlike graph models, it can be intuitively implemented in a relational DBMS. However, several adaptations are still needed for this model to deal with data linkage as in graph models.
 
Graph models, though requiring more specific storage systems such as RDF or graph DBMSs, are still advantageous because they allow to automatically enrich the lake with information that facilitate and enhance future analyses. 
Nevertheless, the three subcategories of graph models need to be all integrated together for this purpose. This remains an open issue because at most two of these graph approaches are simultaneously implemented in metadata systems from the literature \cite{Diamantini2018,Halevy2016B}. The MEDAL metadata model does include all three subcategories of graph models~\cite{Sawadogo2019B}, but is not implemented yet.

\subsection{Metadata Generation}
\label{sec:metadata.generation}  

Most of the data ingestion tools from Section~\ref{sec:archi.techno.ingest} can also serve to extract metadata. For instance, Suriarachchi and Plale use Apache Flume to retrieve data provenance in a data lake \cite{Suriarachchi2016}. Similarly, properties and semantic metadata can be obtained through specialized protocols such as the Comprehensive Knowledge Archive Network (CKAN), an open data storage management system~\cite{Terrizzano2015}.

A second kind of technologies is more specific to metadata generation. For instance, Apache Tika helps  detect the MIME type and language of objects \cite{Quix2016}. Other tools such as Open Calais and IBM's Alchemy API can also enrich data 
through inherent entity identification, relationship inference and event detection~\cite{Farid2016}.

Ad-hoc algorithms can also generate metadata. For example, Singh et al. show that Bayesian models allow detecting links between data attributes \cite{Singh2016}. Similarly, several authors propose algorithms to discover schemas or constraints in semi-structured data \cite{Beheshti2017,Klettke2017,Quix2016}. 

Last but not least, Apache Atlas~\cite{ApacheAtlas}, a widely used metadata framework~\cite{Russom2017}, features advanced metadata generation methods through so-called hooks, which are native or custom scripts that rely on logs to generate metadata. Hooks notably help Atlas automatically extract lineage metadata and propagate tags on all derivations of tagged data.


\subsection{Features of Data Lake Metadata Systems}
\label{sec:metadata.features}  

A data lake made inoperable by lack of proper metadata management is called a data swamp \cite{Khine2017}, data dump \cite{Inmon2016,Suriarachchi2016} or one-way data lake \cite{Inmon2016}, with data swamp being the most common terms. In such a flawed data lake, data are ingested, but can never be extracted. Thus, a data swamp is unable to ensure any analysis. 
Yet, to the best of our knowledge, there is no objective way to measure or compare the efficiency  of data lake metadata systems. Therefore, we first introduce in this section 
a list of expected features for a metadata system. Then, we present a comparison of eighteen data lake metadata systems with respect to these features.   

\subsubsection{Feature Identification} 
\label{sec:metadata.features.identif}  

Sawadogo et al. identify six features that a data lake metadata system should ideally implement to be considered comprehensive~\cite{Sawadogo2019B}.

\begin{enumerate}
\item \textbf{Semantic Enrichment (SE)} is also known as semantic annotation \cite{Hai2016} or semantic profiling \cite{Ansari2018}. It involves adding information such as title, tags, description and more to make the data comprehensible \cite{Terrizzano2015}. This is commonly done using knowledge bases such as ontologies \cite{Ansari2018}.
Semantic annotation plays a vital role in data lakes, since it makes the data meaningful by providing informative summaries \cite{Ansari2018}. In addition, semantic metadata could be the basis of link generation between data \cite{Quix2016}. For instance, data objects with the same tags could be considered linked.  


\item \textbf{Data Indexing (DI)} is commonly used in the information retrieval and database domains to quickly find a data object. Data indexing is done by building and enriching some data structure that enables efficient data retrieval from the lake. Indexing can serve for both simple keyword-based retrieval and more complex querying using patterns. All data, whether structured, semi-structured or unstructured, benefit from indexing \cite{Singh2016}. 


\item \textbf{Link Generation (LG)}  consists in identifying and integrating links between lake data. This can be done either by ingesting pre-existing links from data sources or by detecting new links.  
Link generation allows additional analyses. For instance, similarity links can serve to recommend to lake users data close to the data they currently use \cite{Maccioni2018}. In the same line, data links can be used to automatically detect clusters of strongly linked data \cite{Farrugia2016}. 


\item \textbf{Data Polymorphism (DP)} is the simultaneous management of several data representations in the lake. A data representation 
of, e.g., a textual document, may be a tag cloud or a vector of term frequencies. 
Semi-structured and unstructured data need to be at least partially transformed to be automatically processed \cite{Diamantini2018}. Thus, data polymorphism is relevant as it allows to store and reuse transformed data. This makes analyses easier and faster  by avoiding the repetition of certain processes \cite{Stefanowski2017}.


\item \textbf{Data Versioning (DV)} expresses a metadata system's ability to manage update operations, while retaining the previous data states. 
It is very relevant to data lakes, since it ensures process reproducibility and 
the detection and correction of inconsistencies  \cite{Bhattacherjee2018}. Moreover, data versioning allows branching and concurrent data evolution \cite{Hellerstein2017}.


\item \textbf{Usage Tracking (UT)} consists in managing information about user interactions with the lake. Such interactions are commonly creation, read and update operations.
This allows to transparently follow the evolution of data objects. In addition, usage tracking can serve for data security, either by explaining data inconsistencies or through intrusion detection. 
Usage tracking and data versioning are related, since update interactions often induce new data versions. However, they are distinct features as they can be implemented independently~\cite{Beheshti2017,Suriarachchi2016}.
\end{enumerate}

\subsubsection{Metadata System Comparison} 
\label{sec:metadata.features.compar}  
We present in Table~\ref{tab:metadata.features.compar} a comparison of eighteen state-of-the-art metadata systems and models with respect to the features they implement \cite{Sawadogo2019B}. We distinguish metadata models from implementations. Models are indeed quite theoretical and describe the conceptual organization of metadata. In contrast, implementations follow a more operational approach, but are usually little detailed, mainly focusing on a description of the resulting system instead of the applied methodology.
This comparison also considers metadata systems that are not explicitly associated with the concept of data lake by their authors, but whose characteristics allow  to be considered as such, e.g., the Ground metadata model \cite{Hellerstein2017}.

The comparison shows that the most comprehensive metadata system with respect to the features we propose is MEDAL, with all features covered. However, it is not implemented yet. The next best systems are GOODS and CoreKG, with five out of six features implemented. However, they are black box metadata systems, with few details on metadata conceptual organization. Thus, the Ground metadata model may be preferred, since it is much more detailed and almost as complete (four out of six features).

Eventually, two of the six features defined in Section~\ref{sec:metadata.features.identif} may be considered advanced.  Data polymorphism and data versioning are indeed mainly found in the most complete systems such as GOODS, CoreKG and Ground. Their absence from most of metadata systems can thus be attributed to implementation complexity.

  \begin{table*}[ht]
            \centering
           
            \begin{tabular}{r c c c c c c c c}
            \hline 
                \textbf{System} &  \textbf{Type} & \textbf{SE} & \textbf{DI} & \textbf{LG} & \textbf{DP}  & \textbf{DV} & \textbf{UT}\\  \hline  
               
                SPAR~\cite{Fauduet2010} & \multirow{1}{0.5cm}{$\blacklozenge \sharp$} & \multirow{1}{0.5cm}{\checkmark} & \multirow{1}{0.5cm}{\checkmark} & \multirow{1}{0.5cm}{\checkmark} & 
                \multirow{1}{0.5cm}{} & 
                \multirow{1}{0.5cm}{} & 
                \multirow{1}{0.5cm}{\checkmark}\\ 
                \hline

                \cite{Alrehamy2015} & \multirow{1}{0.5cm}{$\blacklozenge$} & \multirow{1}{0.5cm}{\checkmark} & 
                \multirow{1}{0.5cm}{} & 
                \multirow{1}{0.5cm}{\checkmark} & 
                \multirow{1}{0.5cm}{} & 
                \multirow{1}{0.5cm}{} & 
                \multirow{1}{0.5cm}{}\\ 
                \hline

               \cite{Terrizzano2015}& \multirow{1}{0.5cm}{$\blacklozenge$} & \multirow{1}{0.5cm}{\checkmark} & \multirow{1}{0.5cm}{\checkmark} & 
               \multirow{1}{0.5cm}{} & 
               \multirow{1}{0.5cm}{} & 
               \multirow{1}{0.5cm}{\checkmark} & \multirow{1}{0.5cm}{\checkmark}\\ 
               \hline

                Constance~\cite{Hai2016}& \multirow{1}{0.5cm}{$\blacklozenge$} & \multirow{1}{0.5cm}{\checkmark} & \multirow{1}{0.5cm}{\checkmark} & 
                \multirow{1}{0.5cm}{} & 
                \multirow{1}{0.5cm}{} & 
                \multirow{1}{0.5cm}{} & 
                \multirow{1}{0.5cm}{}\\ 
                \hline

                GEMMS~\cite{Quix2016} & 
                \multirow{1}{0.5cm}{$\lozenge$} & \multirow{1}{0.5cm}{\checkmark} & 
                \multirow{1}{0.5cm}{} & 
                \multirow{1}{0.5cm}{} & 
                \multirow{1}{0.5cm}{} & 
                \multirow{1}{0.5cm}{} & 
                \multirow{1}{0.5cm}{}\\ 
                \hline

                CLAMS~\cite{Farid2016} & \multirow{1}{0.5cm}{$\blacklozenge$} & \multirow{1}{0.5cm}{\checkmark} & 
                \multirow{1}{0.5cm}{} & 
                \multirow{1}{0.5cm}{} & 
                \multirow{1}{0.5cm}{} & 
                \multirow{1}{0.5cm}{} & 
                \multirow{1}{0.5cm}{}\\ 
                \hline

                \cite{Suriarachchi2016} & 
                \multirow{1}{0.5cm}{$\lozenge$} & 
                \multirow{1}{0.5cm}{} & 
                \multirow{1}{0.5cm}{} & 
                \multirow{1}{0.5cm}{} & 
                \multirow{1}{0.5cm}{\checkmark} & 
                \multirow{1}{0.5cm}{} & 
                \multirow{1}{0.5cm}{\checkmark}\\ 
                \hline

                \cite{Singh2016} & 
                \multirow{1}{0.5cm}{$\blacklozenge$} & \multirow{1}{0.5cm}{\checkmark} & \multirow{1}{0.5cm}{\checkmark} & \multirow{1}{0.5cm}{\checkmark} & \multirow{1}{0.5cm}{\checkmark} & 
                \multirow{1}{0.5cm}{} & 
                \multirow{1}{0.5cm}{}\\ \hline

                \cite{Farrugia2016} & \multirow{1}{0.5cm}{$\blacklozenge$} & \multirow{1}{0.5cm}{} & 
                \multirow{1}{0.5cm}{} & 
                \multirow{1}{0.5cm}{\checkmark} & 
                \multirow{1}{0.5cm}{} & 
                \multirow{1}{0.5cm}{} & 
                \multirow{1}{0.5cm}{}\\ 
                \hline

                GOODS~\cite{Halevy2016B}& \multirow{1}{0.5cm}{$\blacklozenge$} & \multirow{1}{0.5cm}{\checkmark} & \multirow{1}{0.5cm}{\checkmark} & \multirow{1}{0.5cm}{\checkmark} & 
                \multirow{1}{0.5cm}{} & 
                \multirow{1}{0.5cm}{\checkmark} & \multirow{1}{0.5cm}{\checkmark}\\ 
                \hline

                CoreDB~\cite{Beheshti2017}& \multirow{1}{0.5cm}{$\blacklozenge$} & \multirow{1}{0.5cm}{} & 
                \multirow{1}{0.5cm}{\checkmark} & 
                \multirow{1}{0.5cm}{} & 
                \multirow{1}{0.5cm}{} & 
                \multirow{1}{0.5cm}{} & 
                \multirow{1}{0.5cm}{\checkmark}\\ 
                \hline

                Ground~\cite{Hellerstein2017}& \multirow{1}{0.5cm}{$\lozenge \sharp$} & \multirow{1}{0.5cm}{\checkmark} & \multirow{1}{0.5cm}{\checkmark} & 
                \multirow{1}{0.5cm}{} & 
                \multirow{1}{0.5cm}{} & 
                \multirow{1}{0.5cm}{\checkmark} & \multirow{1}{0.5cm}{\checkmark}\\ 
                \hline

                KAYAK~\cite{Maccioni2018}& \multirow{1}{0.5cm}{$\blacklozenge$} & \multirow{1}{0.5cm}{\checkmark} & \multirow{1}{0.5cm}{\checkmark} & \multirow{1}{0.5cm}{\checkmark} & 
                \multirow{1}{0.5cm}{} & 
                \multirow{1}{0.5cm}{} & 
                \multirow{1}{0.5cm}{}\\ 
                \hline

                CoreKG~\cite{Beheshti2018}& \multirow{1}{0.5cm}{$\blacklozenge$} & \multirow{1}{0.5cm}{\checkmark} & \multirow{1}{0.5cm}{\checkmark} & \multirow{1}{0.5cm}{\checkmark} & \multirow{1}{0.5cm}{\checkmark} & 
                \multirow{1}{0.5cm}{} & 
                \multirow{1}{0.5cm}{\checkmark}\\ 
                \hline
                
                \cite{Diamantini2018}& 
                \multirow{1}{0.5cm}{$\lozenge$} & \multirow{1}{0.5cm}{\checkmark} & 
                \multirow{1}{0.5cm}{} & 
                \multirow{1}{0.5cm}{\checkmark} & \multirow{1}{0.5cm}{\checkmark} & 
                \multirow{1}{0.5cm}{} & \multirow{1}{0.5cm}{}\\ 
                \hline
                
                 \cite{Mehmood2019}& \multirow{1}{0.5cm}{$\blacklozenge$} & \multirow{1}{0.5cm}{\checkmark} & \multirow{1}{0.5cm}{\checkmark} & \multirow{1}{0.5cm}{} & \multirow{1}{0.5cm}{} & 
                 \multirow{1}{0.5cm}{} & 
                 \multirow{1}{0.5cm}{}\\ \hline
                
                 CODAL \cite{Sawadogo2019A}& \multirow{1}{0.5cm}{$\blacklozenge$} & \multirow{1}{0.5cm}{\checkmark} & \multirow{1}{0.5cm}{\checkmark} & \multirow{1}{0.5cm}{\checkmark} & \multirow{1}{0.5cm}{\checkmark} & 
                 \multirow{1}{0.5cm}{} & 
                 \multirow{1}{0.5cm}{}\\ \hline

                 MEDAL \cite{Sawadogo2019B}& \multirow{1}{0.5cm}{$\lozenge$} & \multirow{1}{0.5cm}{\checkmark} & \multirow{1}{0.5cm}{\checkmark} & \multirow{1}{0.5cm}{\checkmark} & \multirow{1}{0.5cm}{\checkmark} & \multirow{1}{0.5cm}{\checkmark} & \multirow{1}{0.5cm}{\checkmark}\\ \hline
            \end{tabular}
            
            \begin{flushleft}
                $\blacklozenge:~$Implementations~~$\lozenge:~$Metadata models\\
                $\sharp \ :~$Model or implementation akin to a data lake
            \end{flushleft}
             
            
            \caption{Comparison of data lake metadata systems \cite{Sawadogo2019B}}
            \label{tab:metadata.features.compar}
        \end{table*}

\section{Pros and Cons of Data Lakes}
\label{sec:proscons}

In this section, we account for the benefits of using a data lake instead of more traditional data management systems, but also identify the pitfalls that may correspond to these expected benefits.

An important motivating feature in data lakes is \textbf{cheap storage}. Data lakes are ten to one hundred times less expensive to deploy than traditional decision-oriented databases. This can be attributed to the usage of open-source technologies such as HDFS \cite{Khine2017,Stein2014}. Another reason is that the cloud storage often used to build data lakes reduces the cost of storage technologies. That is, the data lake owner pays only for actually used resources. However, the use of HDFS may still fuel \textbf{misconceptions}, with the concept of data lake remaining ambiguous for many potential users. It is indeed often considered either as a synonym or a marketing label closely related to the HDFS technology \cite{Alrehamy2015,Grosser2016}.

Another feature that lies at the core of the data lake concept is \textbf{data fidelity}. Unlike in traditional decision-oriented databases, original data are indeed preserved in a data lake to avoid any data loss that could occur from data preprocessing and transformation operations \cite{Ganore2015,Stein2014}. Yet, data fidelity induces a high risk of \textbf{data inconsistency} in data lakes, due to data integration from multiple, disparate sources without any transformation \cite{OLeary2014}. 

One of the main benefits of data lakes is that they allow exploiting and analyzing \textbf{unstructured data} \cite{Ganore2015,Laskowski2016,Stein2014}. This is a significant advantage when dealing with big data, which are predominantly unstructured \cite{Miloslavskaya2016}. Moreover, due to the schema-on-read approach, data lakes can comply with any data type and format~\cite{Cha2018,Ganore2015,Khine2017,Madera2016}. Thence, data lakes enable a wider range of analyses than traditional decision-oriented databases, i.e., data warehouses and datamarts, and thus show better  \textbf{flexibility and agility}. 
However, although the concept of data lake dates back from 2010, it has only been put in practice in the mid-2010's. Thus, implementations vary, are still maturing and there is a \textbf{lack of methodological and technical standards}, which sustains confusions about data lakes. Finally, due to  the absence of an explicit schema, \textbf{data access services} and  APIs are essential to enable knowledge extraction in a data lake. In other words, a data access service is a must to successfully build a data lake \cite{Alrehamy2015,Inmon2016}, while such a service is not always present.

Next, an acclaimed advantage of data lakes over data warehouses is \textbf{real-time data ingestion}. Data are indeed ingested in data lakes without any transformation, which avoids any time lag between data extraction from sources and their ingestion in the data lake \cite{Ganore2015,Laskowski2016}. But as a consequence, a data lake requires an efficient \textbf{metadata system} for ensuring data access. However, the problem lies in the ``how", i.e., the use of inappropriate methods or technologies to build the metadata system can easily turn the data lake into an inoperable data swamp~\cite{Alrehamy2015}. 

More technically, data lakes and related analyses are typically implemented using distributed technologies, e.g., HDFS, MapReduce, Apache Spark, Elasticsearch, etc. Such technologies usually provide a \textbf{high scalability} \cite{Fang2015,Miloslavskaya2016}. Furthermore, most technologies used in data lakes have replication mechanisms, e.g., Elasticsearch, HDFS, etc. Such technologies allow a high resilience to both hardware and software failure and enforce \textbf{fault tolerance} \cite{John2017}.

Eventually, data lakes are often viewed as sandboxes where analysts can ``play", i.e., access and prepare data so as to perform various, specific, \textbf{on-the-fly analyses} \cite{Russom2017,Stein2014}. However, such a scenario requires \textbf{expertise}. Data lake users are indeed typically data scientists \cite{Khine2017,Madera2016}, which contrasts with traditional decision systems, where business users are able to operate the system. Thus, a data lake induces a greater need for specific, and therefore more expensive, profiles. Data scientists must indeed master a wide knowledge and panoply of technologies. 

Moreover, with the integration in data lakes of structured, semi-structured and unstructured, expert data scientists can discover links and \textbf{correlations between heterogeneous data} \cite{Ganore2015}. Data lakes also allow to easily integrate data ``as is" from external sources, e.g., the Web or social media. Such external data can then be associated with proprietary data to generate new knowledge through \textbf{cross-analyses} \cite{Laskowski2016}. However, several statistical and Artificial Intelligence (AI) approaches are not originally designed for parallel operations, nor for streaming data, e.g., K-means or K-Nearest Neighbors. Therefore, it is necessary to \textbf{readjust classical statistical and AI approaches} to match the distributed environments often used in data lakes \cite{OLeary2014}, which sometimes proves difficult.

\section{Conclusion}
\label{sec:conclusion}

\begin{figure}[hbt]
\centering
\includegraphics[width=12cm]{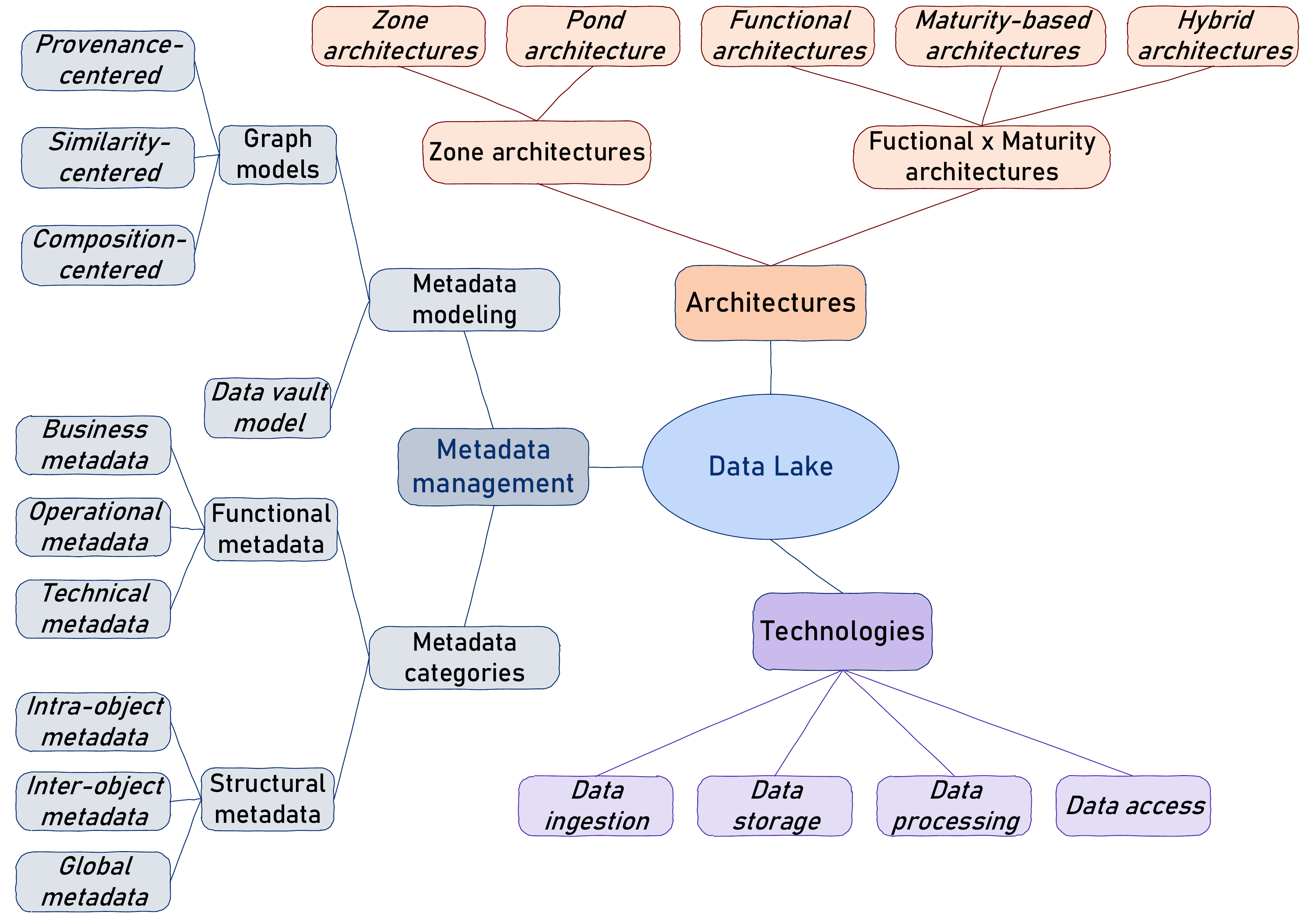} 
\caption{Key concepts investigated in this survey}
\label{fig:maintopics}
\end{figure}

In this survey paper, we establish a comprehensive state of the art of the different approaches to design, and conceptually build a data lake. First, we state the definitions of the data lake concept and complement the best existing one. Second, we investigate alternative architectures and technologies for data lakes, and propose a new typology of data lake architectures. Third, we review and discuss the metadata management techniques used in data lakes. We notably classify metadata and introduce the features that are necessary to achieve a full metadata system. Fourth, we discuss the pros and cons of data lakes. Fifth, we summarize by a mind map the key concepts introduced in this paper (Figure~\ref{fig:maintopics}).

Eventually, in echo to the topics we chose \textit{not} to address in this paper (Section~\ref{sec:intro}), we would like to open the discussion on important current research issues in the field of data lakes. 

\textbf{Data integration and transformation} have long been recurring issues. Though delayed, they are still present in data lakes and made even more challenging by big data volume, variety, velocity and lack of veracity. Moreover, when transforming such data, User-Defined Functions (UDFs) must very often be used (MapReduce tasks, typically). In ETL and ELT processes, UDFs are much more difficult to optimize than classical queries, an issue that is not addressed yet by the literature \cite{Stefanowski2017}.
    
With data storage solutions currently going beyond HDFS in data lakes, \textbf{data interrogation} through metadata is still a challenge. Multistores and polystores indeed provide unified solutions for structured and semi-structured data, but do not address unstructured data, which are currently queried separately through index stores. Moreover, when considering data gravity \cite{Madera2016}, virtual data integration becomes a relevant solution. 
    Yet, mediation approaches are likely to require new, big data-tailored query optimization and caching approaches \cite{Quix2018,Stefanowski2017}.
    
\textbf{Unstructured datasets}, although unanimously acknowledged as ubiquitous and sources of crucial information, are very little specifically addressed in data lake-related literature. Index storage and text mining are usually mentioned, but there is no deep thinking about global querying or analysis solutions. Moreover, exploiting other types of unstructured data but text, e.g., images, sounds and videos, is not even envisaged as of today.
    
Again, although all actors in the data lake domain stress the importance of \textbf{data governance} to avoid a data lake turning into a data swamp, data quality, security, life cycle management and metadata lineage are viewed as risks rather than issues to address \textit{a priori} in data lakes \cite{Madera2016}. Data governance principles are indeed currently seldom turned into actual solutions.
    
Finally, \textbf{data security} is currently addressed from a technical point of view in data lakes, i.e., through access and privilege control, network isolation, e.g., with Docker tools \cite{Cha2018}, data encryption and secure search engines \cite{Maroto2018}. However, beyond these issues and those already addressed by data governance (integrity, consistency, availability) and/or related to the European General Data Protection Regulation (GDPR), by storing and cross-analyzing large volumes of various data, data lakes allow mashups that potentially induce serious breaches of data privacy~\cite{Joss2016}. Such issues are still researched as of today.


%
%

\bibliographystyle{spmpsci}      
\bibliography{biblio}   

%
%

\end{document}